\documentclass[aps, prl, amsmath, amssymb, reprint,
superscriptaddress, showkeys]{revtex4-1}
\pdfoutput=1

\usepackage{amsfonts,amssymb,amsmath,graphicx}
\usepackage{color}
\usepackage{multirow}
\usepackage{siunitx} %for the angle symbol \ang
\usepackage{bm} % bold symbols in formulas
\usepackage{xcolor}

\usepackage[pdftex,
            ]{hyperref}

\def\?#1{}

\newcommand{\commentout}[1]{}
\newcommand{\comment}[1]{}

\usepackage{bbm} %1-Matrix

\def\be{\begin{equation}}
\def\ee{\end{equation}}
\def\ber{\begin{eqnarray}}
\def\eer{\end{eqnarray}}

\begin{document}

\title{Magnetoconductance, Quantum Hall Effect, and Coulomb Blockade \\ 
       in Topological Insulator Nanocones}
\author{Raphael Kozlovsky, Ansgar Graf, Denis Kochan, Klaus Richter, Cosimo Gorini}
\email[]{cosimo.gorini@physik.uni-regensburg.de}
\affiliation{Institut f{\"u}r Theoretische Physik, Universit{\"a}t Regensburg, 
	93040 Regensburg, Germany}

\date{\today}

\begin{abstract}

Magnetotransport through cylindrical topological insulator (TI) nanowires is governed by the interplay between quantum confinement and geometric (Aharonov-Bohm and Berry) phases. 
Here, we argue that the much broader class of TI nanowires with varying radius $\textendash$ for which a \textit{homogeneous} coaxial magnetic field induces a \textit{varying} Aharonov-Bohm flux that gives rise to a non-trivial mass-like potential along the wire $\textendash$ is accessible by studying its simplest member, a TI nanocone. 
%Here, we approach the much broader class of topological insulator nanowires with varying radius $\textendash$ for which a \textit{homogeneous} coaxial magnetic field induces a \textit{varying} Aharonov-Bohm flux that gives rise to a non-trivial mass-like potential along the wire $\textendash$ by studying its simplest member, a topological insulator nanocone.
Such nanocones allow to observe intriguing mesoscopic transport phenomena: While the conductance in a perpendicular magnetic field is quantized due to higher-order topological hinge states, it shows resonant transmission through Dirac Landau levels in a coaxial magnetic field. Furthermore, it may act as a quantum magnetic bottle, confining surface Dirac electrons and leading to a largely interaction-dominated regime of Coulomb blockade type. We show numerically that the above-mentioned effects occur for experimentally accessible values of system size and magnetic field, suggesting that TI nanocone junctions may serve as building blocks for Dirac electron optics setups.

\end{abstract}

\maketitle

\begin{figure}
	\includegraphics[width=1\columnwidth]{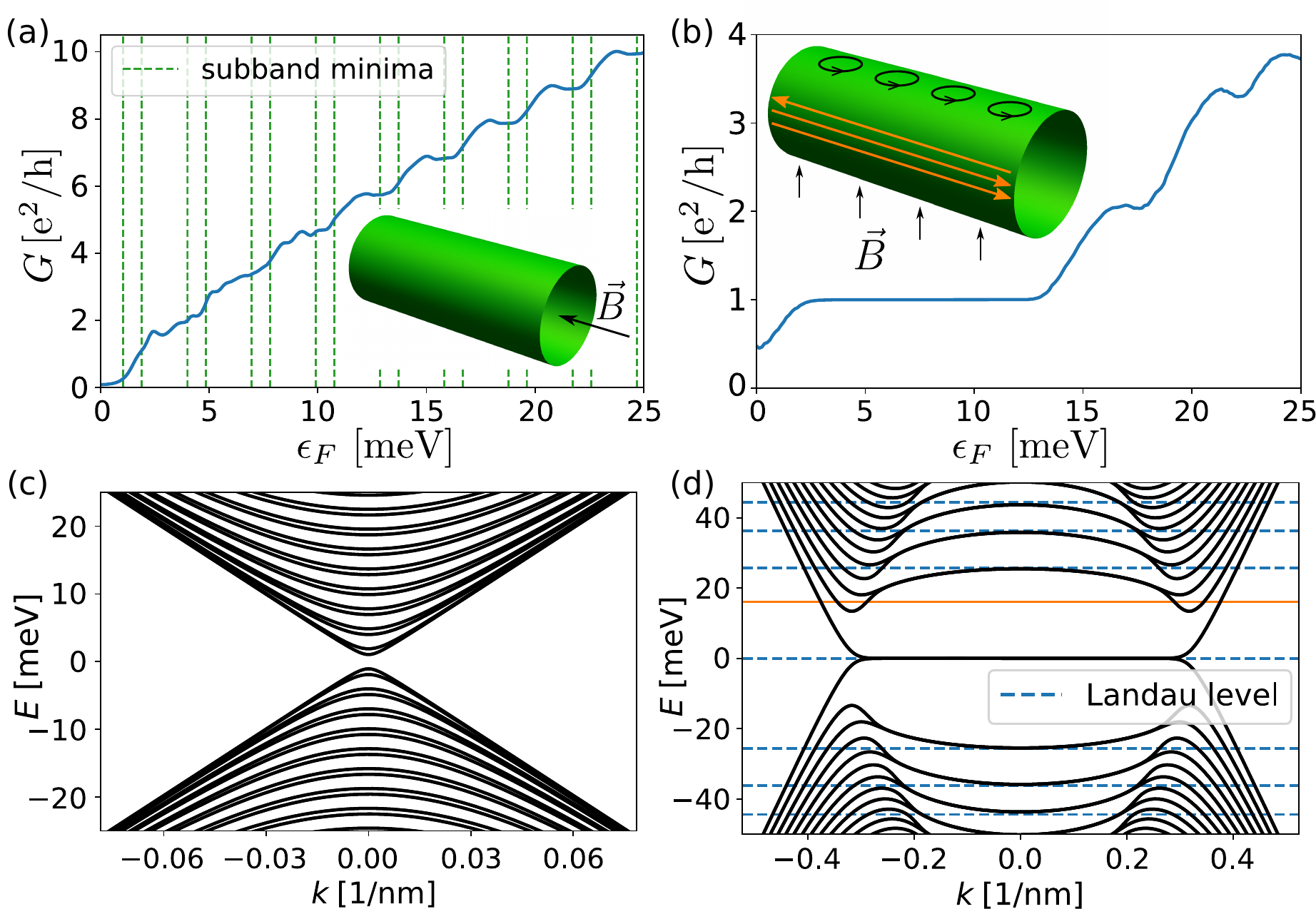}
	\caption{
	Disorder-averaged conductance as function of the Fermi energy $\epsilon_F$ (top panels) and band structure (with longitudinal wave number $k$, bottom)
        of a cylindrical TI nanowire in a coaxial (left) and perpendicular (right)
        magnetic field $B\!=\!2\,$T.
        The orange arrows in (b) sketch 1D hinge states at an energy marked with an orange line in (d).
	Wire size: length $L\!=\!600\,$nm, circumference $C\!=\!700\,$nm. We use a
	Gaussian-correlated disorder average taken over 600 configurations, and a Fermi velocity of $v_F=\SI{5e5}{m/s}$ here and throughout. 
       % Average based on Gaussian disorder is taken over 600 configurations.
        %, for details see the Supp. Mat. \cite{SuppMat}. 
         }
	\label{fig_cylinder}
\end{figure}

Electronic transport across phase-coherent structures has been a central topic of solid state research
ever since the birth of mesoscopic physics some 40 years ago.
While the complexity of mesoscopic setups has steadily increased, from the simple gate-defined quantum point
contacts of the '80s \cite{vanwees1988} to elaborate present-day electron optics circuits in semiconductors 
\cite{bocquillon2013}  and graphene \cite{kumada2016,makk2018},
their structure remains in the vast majority of cases planar -- {\it i.e}.~transport takes place
in flat two-dimensional (2D) space.
Exceptions to the 2D scenario are samples based on carbon nanotubes and 3D topological insulator (3DTI) 
nanowires \cite{peng2009,dufouleur2013,cho2015,ziegler2018}. 
3DTIs are bulk band insulators hosting protected 2D surface metallic states \`a la Dirac \cite{hasan2010}.
In mesoscopic nanostructures built out of 3DTIs low-temperature phase-coherent transport 
takes place on a 2D Dirac metal wrapped around an insulating 3D bulk.  
As such, it is strongly dependent on a peculiar conjunction of structural (real space)
and spectral (reciprocal space) geometrical properties.
This has remarkable consequences for a topological insulator nanowire (TINW) with constant circular cross section in a coaxial magnetic field, shown in Fig.~\ref{fig_cylinder}(a). Its magnetoconductance reflects a non-trivial interplay between two fundamentals of mesoscopic physics: quantum confinement and geometric [Aharonov-Bohm (AB) and Berry] phases
\cite{ostrovsky2010,bardarson2010,
dufouleur2013,cho2015,ziegler2018,dufouleur2018}. 

\begin{figure}
	\includegraphics[width=.9\columnwidth]{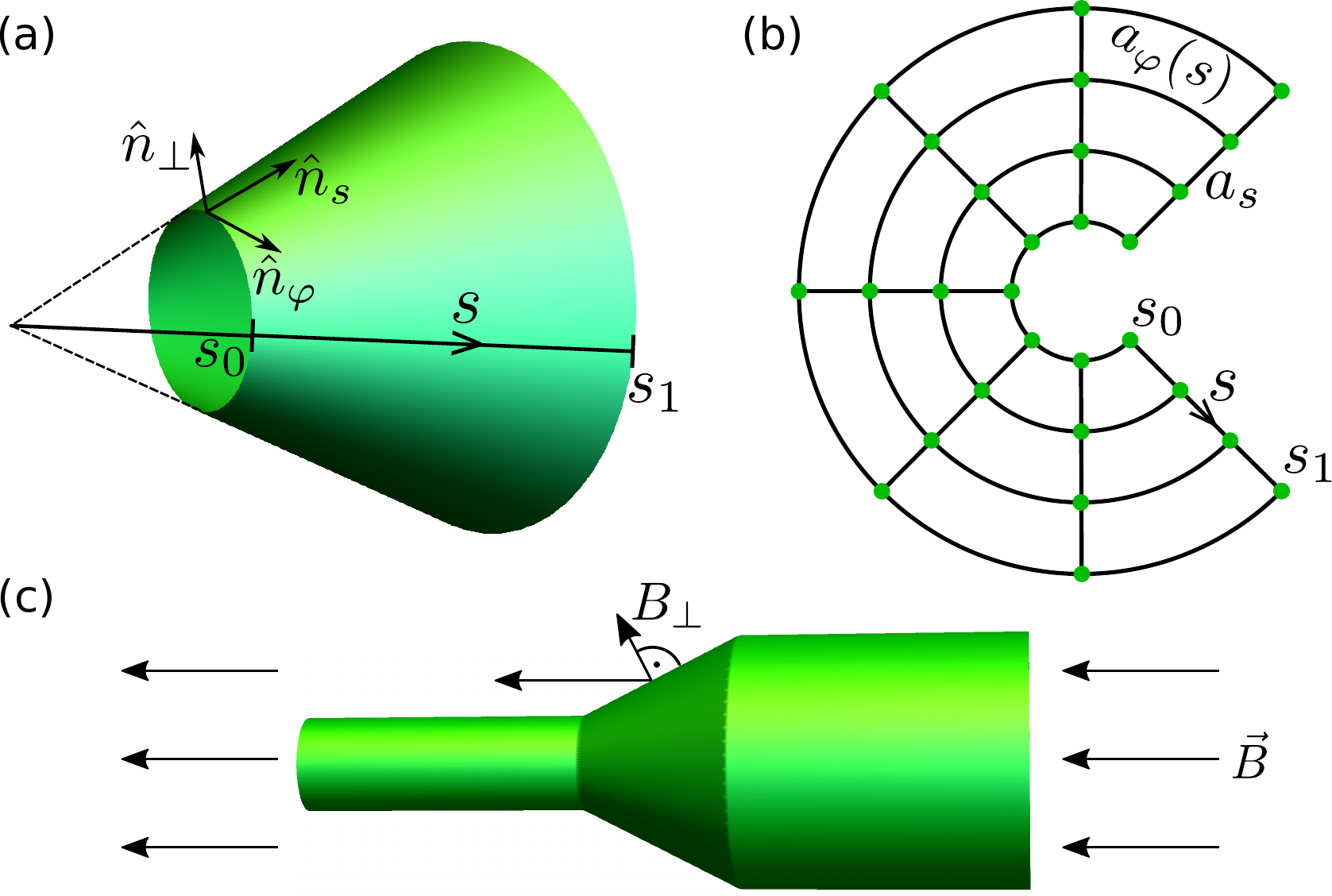}  % was 0.9 #
	\caption{(a) Sketch of a TINC extended to its conical singularity with  
        local coordinate vectors $ \hat{n}_\varphi, \hat{n}_s,\hat{n}_\perp$. 
       (b) Unfolded cone grid used for numerical simulations.
       (c) TINC in coaxial B-field (with $B_\perp=\bm{B}\cdot \hat{n}_\perp$)  with cylindrical leads attached.%, acting as wave guides in magnetotransport simulations. 
        }
	\label{fig_grid_cone}
\end{figure}

An interesting twist offered by 3DTIs
is the possibility of engineering shaped TINWs with a variable cross section, 
{\it e.g}.~truncated TI nanocones (TINCs) as sketched in Fig.~\ref{fig_grid_cone}(a).
When a coaxial magnetic field is switched on, shaped TINWs possess the unique feature that surface charge carriers traversing the wire, experience not only a variation of the centrifugal potential, but also a spatially changing AB flux -- a property that cannot easily be realized with bulk conductors.
%These possess the unique feature that surface Dirac charge carriers, while traversing the shaped TINW in a coaxial magnetic field, experience not only the variation of the centrifugal potential, a spatially changing AB flux (a property that cannot easily be realized with bulk conductors).
%This gives rise to a flux-induced effective mass potential along the TINC which, in contrast to cylindrical TINWs, does not possess any magnetic field periodicity. 
%Together with Dirac electron physics this implies a variety of interesting 
%mesoscopic transport phenomena, including resonant transport through Dirac Landau levels and
%magnetically induced Coulomb blockade physics. 
This gives rise to a flux-dependent effective mass potential along the TINC, which induces a variety of interesting mesoscopic transport phenomena, including resonant transport through Dirac Landau levels and magnetically induced Coulomb blockade physics.
Due to the non-trivial real space geometry, all such transport regimes can be accessed simply by applying
and tuning a homogeneous magnetic field. 
%%%%%%%%%%%%%%%%%%%%%%%%%

\textit{Cylindrical TINW in a magnetic field} --
For later reference consider cylindrical TINWs of radius $R$
in coaxial \cite{ostrovsky2010,bardarson2010,dufouleur2013,cho2015,ziegler2018,dufouleur2018} 
or perpendicular \cite{Ajiki1993, Vafek2011, Zhang2012, Sitte2012, Brey2014, DeJuan2014, Konig2014, Xypakis2017} 
magnetic fields.
They can be described by the 2D surface Dirac Hamiltonian \cite{zhang2010}
%\be
${H} \!=\!  v_F \left[\frac{1}{2} \left( {p}_\varphi \sigma_\varphi
 + \sigma_\varphi {p}_\varphi \right)+ \sigma_z  {p}_z \right]$
%\label{hamiltonian_cylinder}
%\ee
with Fermi velocity $v_F$, coordinate along the wire axis $z$, and azimuthal angle $\varphi$. 
Furthermore, $p_\varphi \!=\! \frac{-i\hbar}{R} \partial_\varphi$ and $\sigma_\varphi \!=\! \sigma_y \cos\varphi - \sigma_x \sin\varphi$ with Pauli matrices $\sigma_x$, $\sigma_y$, $\sigma_z$. 
%
%The unitary transformation $U \!=\! \exp(i \sigma_z \varphi /2)$ simplifies
%the Hamiltonian to 
%$H = v_F \left(\sigma_z p_z + \sigma_y p_\varphi\right)$,
%while the Dirac electron wave functions acquire antiperiodic boundary conditions.
The unitary transformation $U \!=\! \exp(i \sigma_z \varphi /2)$ yields $H = v_F \left(\sigma_y p_\varphi + \sigma_z p_z \right)$ and antiperiodic boundary conditions for the wave functions.
The associated Berry phase \cite{mikitik1999} shifts the angular momentum quantization
condition by $\hbar/2$, yielding a gapped subband spectrum $E_l(p_z)\!=\!\pm v_F \sqrt{p_z^2 + \hbar^2(l + 1/2)^2/R^2}$, 
with orbital angular momentum quantum number $l \in \mathbb{Z}$.
%%%%%%%%%%%%%%%%%%%%%%
%%%%%%%%%%%%%%%%%%%%
%
%By applying a longitudinal magnetic field which corresponds to a magnetic flux of $\Phi_0/2$ through the cross section, the electrons exhibit an additional phase of $\pi$ and thus periodic boundary conditions are restored which leads to a gap closing.
%
For a coaxial $B$-field generating a magnetic flux $\Phi\!=\! \pi R^2 B$ ($B\!=\!|\bf B|$) through the tube
the problem remains separable and reduces to that of an electron in an AB ring -- 
the cylinder cross section -- times free longitudinal motion.  
The discrete spectrum of the ring is periodic in $\Phi_0$ \cite{eckern1995}, 
and is turned into a series of 1D subbands by free longitudinal motion.
% , see  Fig.~\ref{fig_cylinder} c).
%
Due to the Berry phase the bands are gapped for $\Phi \!=\! n \Phi_0$
% , n\in\mathbb{Z}$, 
and gapless for $\Phi \!=\! (n+1/2) \Phi_0, n\in\mathbb{Z}$ \cite{ostrovsky2010,bardarson2010}. 
%The spectrum is two-fold degenerate for integer and half-integer flux, with a non-degenerate gap-crossing mode appearing in the second case.  At $\Phi \!=\! (n+1/2) \Phi_0$, the gap-crossing ``perfectly conducting channel''  is predicted to yield an $e^2/h$ conductance plateau. 
Figure~\ref{fig_cylinder}(c) shows the generic situation with $\Phi \approx 18.86 \, \Phi_0$.
The corresponding conductance is depicted
in Fig.~\ref{fig_cylinder}(a), increasing with energy as more channels open \cite{ziegler2018}. For details of the numerics see below and the Supp.~Mat.~\cite{SuppMat}.
% , with dips close to each subband opening, where van-Hove singularities in the density of states cause 
% enhanced scattering 

If the magnetic field is orthogonal to the nanowire axis, the situation changes drastically.
The resulting band structure, see Fig.~\ref{fig_cylinder}(d), can be understood qualitatively in classical terms \cite{onorato2011}:
The field component $B_\perp = B \cos\varphi$ perpendicular to the surface
varies along the wire circumference, being maximal for $\varphi\!=\!0,\pi$ (cylinder top and bottom) and zero for $\varphi\!=\!\pi/2,\,3\pi/2$
(sides), where its sign changes. Cyclotron orbits [black circles in Fig.~\ref{fig_cylinder}(b)] of opposite handedness thus form 
on the top and bottom surfaces, while ``snaking'' orbits propagating along the sides appear (orange arrows).
Quantum mechanically, the former lead to Landau level (LL) formation, while the latter 
are chiral 1D hinge states crossing the $B_\perp$-induced gap and 
signaling a higher-order TI phase \footnote{See \cite{schindler2018}.  More precisely, we deal here with a second-order TI phase due to the applied magnetic field, and thus \textit{extrinsic} \cite{Sitte2012}, as opposed to \textit{intrinsic} higher-order phases arising from crystal symmetries \cite{geier2018}.}.
The dashed lines in Fig.~\ref{fig_cylinder}(d) mark the LL energies for $B=B_\perp(\varphi=0)$,
%\be
%%E_n = \rm{sgn(n)} v_F \sqrt{2 \hbar e B_\perp |n|} \; , \; n \in \mathbb{Z},
%\ee
\be
E_n = \mathrm{sgn}(n) \frac{\hbar v_F}{l_B} \sqrt{2 |n|}   \; , \; n \in \mathbb{Z} \,,
\label{LL}
\ee
where $l_B=\sqrt{\hbar/(eB)}$ is the magnetic length.
Around $k\!=\!0$, flat bands represent well-formed LLs.  
The upward (downward) bending $n\!=\!0$ subband ensures the existence of a robust conductance plateau
$G\!=\!e^2/h$ in a large energy window, see Fig.~\ref{fig_cylinder}(b), 
where only one chiral hinge state per side is present, preventing backscattering. The dip at $\epsilon_F=0$ is treated in Ref. \cite{Xypakis2017}.
%
%\vspace{1cm}
%\subsection{cone Hamiltonian, volume trafo, implementation in kwant}  
%- switching to conical geometry \ra expect novel quantum transport phenomena, especially in magnetic field

%%%%%%%%%%%%%%%%%%%%%%%%%
 
\textit{TINC in a coaxial magnetic field} --
We generalize our considerations to a TINC, see Fig.~\ref{fig_grid_cone}(a), a representative building block of TINWs with varying cross section. We parametrize its surface by the azimuthal angle $\varphi$ and the distance $s$ to the 
conical singularity.
The radius is $R(s) \!=\! s\,  \sin(\beta/2) $, $\beta$ denoting 
the cone opening angle and $s_0 \! \! \leq s \! \leq s_1$.
Using standard techniques \cite{Xypakis} and 
after the unitary transformation $U \!=\! \exp(i \sigma_z \varphi /2)$,
% , i.~e.~wave functions satisfy antiperiodic boundary conditions, 
the surface Dirac Hamiltonian reads 
%\footnote{See Ref.~\cite{Xypakis} for a detailed derivation.  
%The Hamiltonian  can most straightforwardly be derived by considering the Dirac operator for 
%spin-1/2 particles on a conical surface, and computing the appropriate spin connection entering 
%the covariant derivative. Details of this procedure can be found in Ref.~\cite{Fecko}.}
\be
{H} = v_F \left[ \left({p}_s - \frac{i \hbar}{2s}\right)  \sigma_z + {p}_\varphi(s) \sigma_y \right] \, ,
\label{H}
\ee
with ${p}_\varphi(s) \!=\! -\frac{i\hbar}{R(s)} \partial_\varphi$.  
%$\hat{H}$ is the Dirac operator for spin-1/2 particles on a space-like surface with conical geometry. Curvy-linear coordinates require to compute the spin connection which enters the covariant derivative of the spinor field. We obtain it by following the conventional procedures discussing the Dirac operator on a general Riemannian manifold, for details see \cite{Fecko} chapters 15 and 22. $\hat{H}$ can also be derived via microscopic considerations as already presented in \cite{Xypakis}.
%with the metric tensor $g_{ij} = [1, R^2(s)]$ 
The spin connection term $-\frac{i\hbar}{2s}$ \cite{Fecko} makes it Hermitian with respect 
to the scalar product 
 \be
 \left< \Psi_1 | \Psi_2 \right> \equiv \int_{s_0}^{s_1} \! \mathrm{d} s \int_{0}^{2\pi} \! \mathrm{d} \varphi \;
  R(s) \Psi_1^*(s, \varphi) \Psi_2(s,\varphi) \, .
 \ee
Transforming $\Psi \rightarrow \widetilde{\Psi}\!=\!  \sqrt{R(s)} \Psi$ 
and $H \rightarrow \tilde{H} \!=\! \sqrt{R(s)} H  (1/\sqrt{R(s)})$ removes the spin connection term from Eq.~\eqref{H} 
% (conformal mapping?) 
and renders the volume form $R(s) {\rm d}\varphi {\rm d} s$ trivial.
This yields
\be
{\tilde{H}} = v_F [ p_s  \sigma_z + p_\varphi(s) \sigma_y ] \, ,
\label{H_tilde}
\ee
with  $\langle  \widetilde{\Psi}_1 | \widetilde{\Psi}_2 \rangle  \! \equiv \! 
 \int \! \mathrm{d} s \int \! \mathrm{d} \varphi 
\widetilde{\Psi}_1^*(s, \varphi) \widetilde{\Psi}_2(s,\varphi)$,
% From now on we will work with the transformed quantities $\tilde{\Psi}, {\tilde H}$.
which is necessary for our numerics \cite{SuppMat}.  

\begin{figure}
	\includegraphics[width=1\columnwidth]{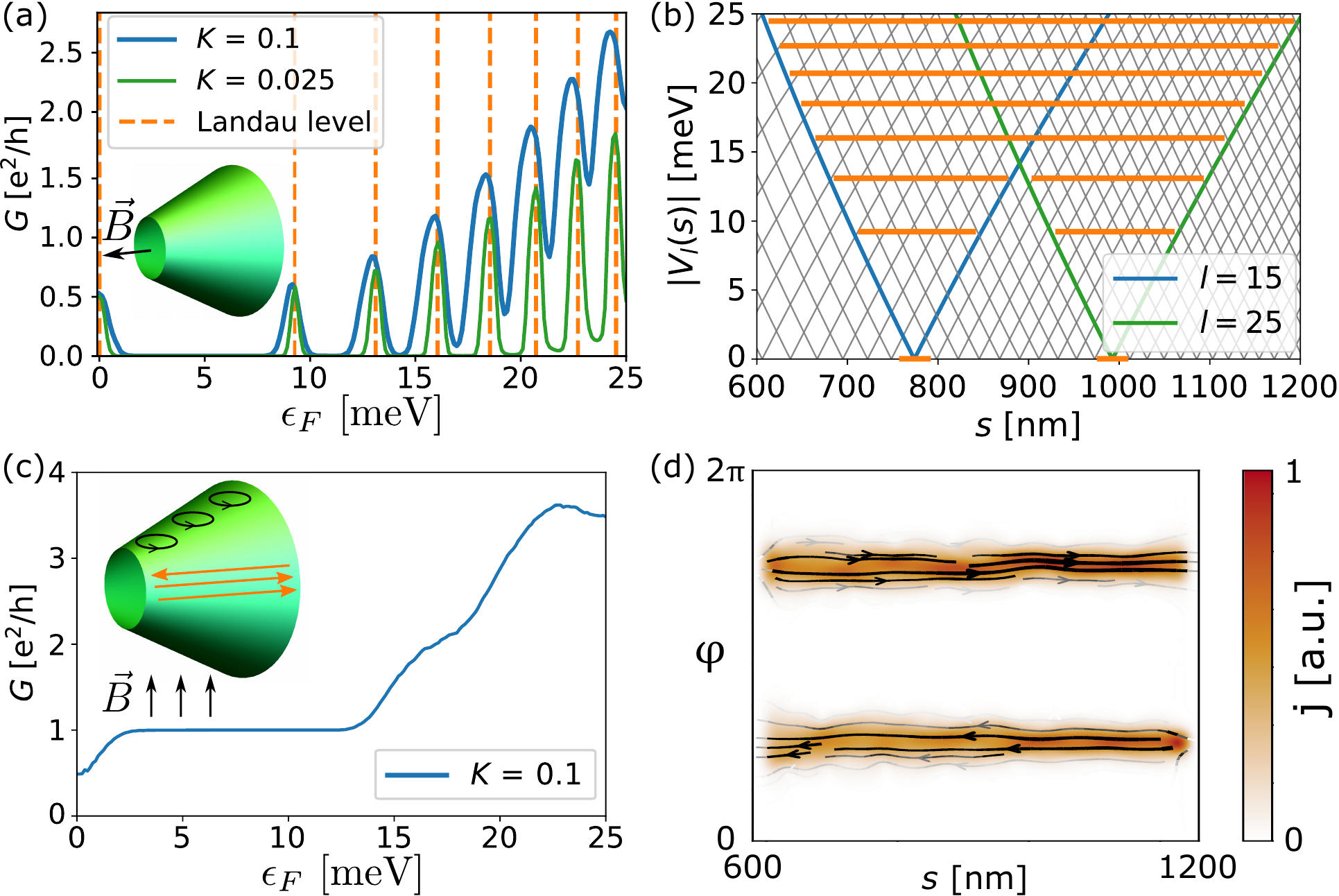}
	\caption{
        Disorder averaged TI nanocone conductance for coaxial and perpendicular magnetic field.
	(a) Coaxial field: resonant transmission through LLs.  
	Stronger disorder (larger $K$ \cite{SuppMat}) broadens and shifts the LLs from their
	unperturbed positions marked by vertical dashed lines \cite{peres2006, dora2008}.
     (b) LL formation in effective wedge potentials.
        %(with energies marked as horizontal lines) is depicted in panel (c) 
        %and explained in the main text.   
        %If energies exceed the effective potentials within the cone, channels fully open for transport and conductance steps 
       % of the order of $e^2/h$, slightly deformed by disorder, appear (not shown).
	(c) Perpendicular field: robust
	conductance plateau due to topologically protected chiral hinge states
	and associated current density in (d). 
        %Step-like features almost completely smoothed by disorder appear for $E \!>\! 13\,{\rm meV}$. 
        Parameters: $B\!=\!2\,$T; cone with (arc) length $600\,$nm
	and opening angle $\beta\!=\!15^{\circ}$, beginning a distance $s_0\!=\!600\,$nm away
	from the conical singularity, yielding minimal and maximal circumferences $C_{\rm min}\!\approx\! 492\,$nm,
	$C_{\rm max}\!\approx \!984\,$nm (see Fig.~\ref{fig_grid_cone}).
        }
	\label{fig_cone}
\end{figure}

In presence of a coaxial magnetic field the problem remains rotationally symmetric and thus separable.
We proceed via the \textit{exact} separation ansatz  
$\widetilde{\Psi}_{nl}(s, \varphi)
\! = \! \mathrm{e}^{i(l+1/2) \varphi} \chi_{nl} (s)$, 
with $ \chi_{nl} (s)$ a two-component spinor. The +1/2 in the exponential originates from the curvature-induced Berry phase, $l$ is the orbital angular momentum quantum number, and
%Since the azimuthal part of $\Psi_{nl}(s,\varphi)$ is anti-periodic, 
%$l \! \in \! \mathbb{Z}$ can still be interpreted as an angular momentum quantum number;
the meaning of 
% the longitudinal quantum number 
$n$ will be clarified shortly. 
Using minimal coupling $p_\varphi(s) \rightarrow p_\varphi(s) \! + \!  e A_\varphi(s)$
yields the 1D  Dirac equation
\be
 \left[v_F {p}_s \sigma_z + V_l(s,B) \sigma_y \right] \chi_{nl}(s) = \epsilon_{nl} \chi_{nl}(s).
\label{1D_Dirac}
\ee
Here, the angular momentum term 
\be
\label{eq_potential}
V_l(s,B) = \frac{v_F\hbar}{R(s)} \left(l + \frac{1}{2} - \frac{\Phi(s, B)}{\Phi_0} \right),
\ee
with $\Phi(s,B) \! = \! \pi R^2(s) B$,
acts as a position-dependent mass potential, and is crucial for predicting the TINC magnetotransport properties -- indeed, 
more generally the properties of arbitrarily shaped, rotationally symmetric TINWs \cite{to_be_published}. 
%as long as rotational symmetry is preserved. 
Its mass-like character becomes evident in Eq.~\eqref{1D_Dirac}, where it couples to $\sigma_y$, while a simple electrostatic potential enters the equation with the identity matrix. Hence, Eq.~\eqref{1D_Dirac} describes 1D Dirac electrons feeling the \textit{effective potential} $|V_l(s)|$.  Dirac electrons Klein-tunnel through electrostatic barriers, but not
through $|V_l(s)|$.
Figure \ref{fig_cone}(b) shows $|V_l(s,B)|$ for a coaxial $B$-field, whose $B_\perp$ yields $l_B \ll s_1-s_0$ for all $l$-values relevant in the presented energy range (only $l=15$ and $l=25$ are colored and labeled, the rest is gray). In transport, an electron injected in mode $l$ feels a distinct effective potential $|V_l(s,B)|$.
For a discussion of the $B$-field dependence of $|V_l(s,B)|$ and the formation of wedges shown in Fig.~\ref{fig_cone}(b) see \cite{SuppMat}.
Note that due to the step-like form of $B_\perp(s)$, cf.~Fig.~\ref{fig_grid_cone}(c), the situation is similar to single-valley graphene subject to a magnetic barrier \cite{demartino2007,masir2008}.
%%%%%%%%%%%%%%%%%%%%%%%%%%%%%%%%%%%%%%%%%55
%%%%%%%%%%%%%%%%%%%%%%%%%%%%%%%%%%%%%%%%%
 
\textit{Magnetotransport through TINCs} --
Our quantum transport simulations including disorder are based on the software package
\textit{kwant} \cite{Groth2014}. 
%In general, direct discretization of a Hamiltonian 
%with a non-trivial volume form such as Eq.~\eqref{H} yields a non-Hermitian tight-binding matrix \cite{SuppMat}.
The tight-binding representation of Eq.~\eqref{H} is non-Hermitian due to the non-trivial volume form \cite{SuppMat}.
The latter is rendered trivial by the transformation ${H}\rightarrow{\tilde H}$, hence Eq.~\eqref{H_tilde} has a Hermitian tight-binding representation.
%However, the above transformation ${H}\rightarrow{\tilde H}$ [see Eq.~\eqref{H_tilde}], provides
%an appropriate basis for the numerical simulations.
Conventional discretization of Eq.~\eqref{H_tilde} leads to the effective lattice shown in Fig.~\ref{fig_grid_cone}(b),
representing the unfolded cone. The transversal lattice constant $a_\varphi(s)$, 
which is part of the hopping integral, is adapted such that $N_\varphi a_\varphi(s) \!=\! U(s)$, 
where $N_\varphi$ is the number of lattice points and $U(s)$ the circumference at position $s$,
and the ends are ``glued'' together (for details see~\cite{SuppMat}).
To compute the conductance, highly-doped semi-infinite cylindrical leads with radii $R(s_0)$ and $R(s_1)$
are attached to the conical scattering region, see Fig.~\ref{fig_grid_cone}(c).

Figure~\ref{fig_cone}(a) shows the TINC conductance for the potential $|V_l(s,B)|$ in Fig.~\ref{fig_cone}(b).
Sharp peaks signal resonant transmission through quasi-bound states of wedge-shaped effective potentials: 
Each wedge (labeled by $l$) hosts a sequence of bound states (labeled by $n$) 
whose energies $\left\{\epsilon_{nl}\right\}$,
marked by horizontal orange lines in Fig.~\ref{fig_cone}(b), are solutions of Eq.~\eqref{1D_Dirac}. 
% Here $l$ labels the wedges and the quantum number $n$ the different bound states within each wedge.
The latter are quantum Hall (QH) states, degenerate in $l$ and forming the Dirac LL given in Eq.~\eqref{LL}.
The LL degeneracy is thus given by the number of wedges within the cone (between 
$s_0$ and $s_1$). 
In a clean TINC degenerate states in adjacent wedges are orthogonal, hence transport is exponentially
suppressed. 
Disorder breaks rotational symmetry
and couples adjacent states leading to broadened resonant tunneling peaks, see Fig.~\ref{fig_cone}(a).
%\commentRK{Indeed, stronger disorder (larger $K$) results in a larger conductance 
%due to the increased coupling between QH states (cf. green and blue curves).}
Indeed, stronger disorder increases the coupling between QH states, 
and thus the conductance (compare green and blue curves). 
Note the close relation to a QH Corbino geometry \cite{rycerz2010}:
Looking at the TINC from the front, its 2D projection is a ring of finite thickness in a homogeneous 
perpendicular magnetic field $B_\perp$. 
%%%%%%%%%%%%%%%%%%%%%%%
\begin{figure}
\includegraphics[width=\columnwidth]{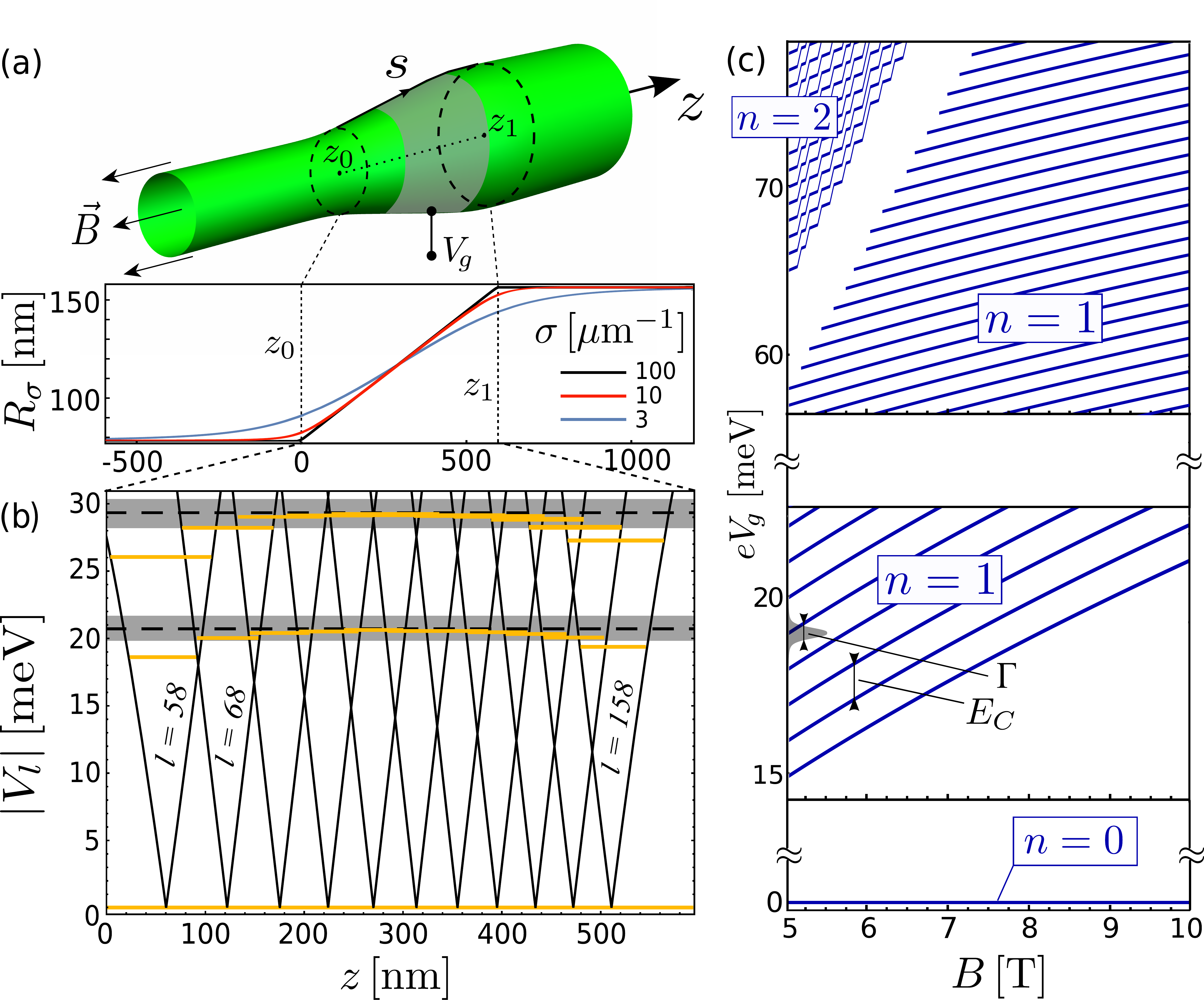}
\caption{Anomalous Coulomb blockade in TI nanocones. 
(a) Smoothed TINC parametrized by $R_\sigma(z)$ \cite{SuppMat} with $\sigma\!=\!10\,\mu$m$^{-1}$ (small $\sigma$ corresponds to strong smoothing, see lower panel). A gate wrapping around the TINC is sketched by the gray region.
(b) $|V_l(z)|$ from $l\!=\!58$ to $l\!=\!158$ (in steps of $\Delta l\!=\!10$) are shown for $B\!=\!10\,$T. QH states (marked by orange lines) with $n\!\neq\!0$ are no longer resonant across the entire junction, leading to tunnel barriers close to $z_0, z_1$. 
%If the lead Fermi energy matches one of the dashed lines marking sharp TINC ($\sigma\!\rightarrow\!\infty$) LL energies, transport is expected to be determined by Coulomb blockade physics. 
Grey shade depicts the disorder broadening $\Gamma$ from Fig.~\ref{fig_cone}(a) for $K=0.1$. 
(c) Addition spectrum of the $n=0, 1, 2$ LLs in the central TINC region, with charging energy $E_C\!=\!1\,$meV.
The $n=1, 2$ LLs, defined by QH states whose energy lies within the disorder broadening,
are Coulomb split by $E_C$, while $n=0$ QH states are not affected by Coulomb repulsion.
%(c) Inner island LL spectrum as function of the magnetic field.  Black lines represent single Dirac QH states, separated by the charging energy $E_c$ and broadenend by $\Gamma$.  
%The large gap at $E\approx 60$ meV for $B=5$ T is the Landau gap between LL $n=1$ (composed of the lower right set of QH states) and $n=2$ (upper left set of QH states).  For increasing $B$ additional QH states join the $n=1$ LL, adding to its degeneracy and thus shifting upwards by $E_c$ the bottom of the $n=2$ LL.
}
\label{fig_CB}
\end{figure}
%%%%%%%%%%%%%%%%%%%%%%

For a TINC in a $B$-field orthogonal to its symmetry axis,
the conductance 
is completely different, see Fig.~\ref{fig_cone}(c).
Its main features -- dip at zero energy, plateau up to $E\! \simeq \!13\,{\rm meV}$ and subsequent
disorder-smoothed steps -- stem from second-order topological hinge states at the sides.
%exhibits completely different 
%features, see Fig.~\ref{fig_cone}(c).
%A dip at zero energy merges into an extended plateau
%reaching $ E\! \simeq \!13\,{\rm meV}$, followed by disorder-smoothed steps at higher energy. 
%As for cylindrical TINW considered before, such features are due to second-order topological hinge states at the sides.
%%Such features are due to second-order topological hinge states at the sides, 
%%similar to the case of the cylindrical TINW considered before.
If $l_B \!\ll\! \pi R(s_0)$, they are indistinguishable from those of a cylindrical TINW, cf.~Figs.~\ref{fig_cylinder}(b) and \ref{fig_cone}(c). 
This is actually true in a much more general sense: As long as top and bottom surface provide enough space for LLs to form, 
the geometry of the TINW in perpendicular B-field is irrelevant for the qualitative conductance features, as opposed to TINWs in coaxial B-field.
The current density associated with the lowest-energy hinge state of the TINC, yielding the robust plateau, 
is plotted in Fig.~\ref{fig_cone}(d) and seen to be chiral -- the current on opposite sides $(\varphi \! = \!
\pi /2, 3\pi/2)$ flows in opposite directions.

The two settings in Fig.~\ref{fig_cone}(a) and (c) correspond to
a longitudinal ($\sigma_{xx}$) and transversal ($\sigma_{xy}$) conductivity
measurement in a conventional 2D QH setup where 
%the measurement of the longitudinal and transverse conductivity,
%$\sigma_{xx}$ and $\sigma_{xy}$, in a conventional 2D QH measurement.
the longitudinal
current density reads $j_x \!=\! \sigma_{xx} E_x\! +\! \sigma_{xy} E_y$.
%$x, y$ denoting the longitudinal and transverse directions. 
Let us adapt this to the TINC.
In a perpendicular magnetic field, $E_x$ vanishes as long as $\epsilon_F$ lies within the plateau
due to lack of backscattering. Hence, the conductance is solely determined by
$\sigma_{xy}$. On the contrary, in a coaxial magnetic field $E_y$ vanishes 
%due to the rotational symmetry of the system,
since metallic states extend across the circumference, 
resulting in a conductance determined by $\sigma_{xx}$ only.

\textit{Coulomb blockade in smoothed TINCs} -- 
Equation \eqref{1D_Dirac} and thus the concept of an effective potential, Eq.~(\ref{eq_potential}), is valid more generally for any rotationally symmetric but arbitrarily shaped TINWs, the coordinate $s$ generalizing to an arc length coordinate \cite{to_be_published}. 
%which is related to the coaxial coordinate $z$ by 
%\be
%s(z)=\int_{z_0}^zdz'\sqrt{1+R'(z)^2}.
%\ee
Exploiting the effective potential, one can devise different wire geometries featuring (tunable) magnetic barriers able to confine Dirac electrons \cite{to_be_published}.  As a simple and paradigmatic example, we consider a smoother, 
more realistically shaped TINC, see Fig.~\ref{fig_CB}(a).  In the regions of smaller slope (close to the leads) $B_\perp$ decreases to zero,
ergo the QH states drop in energy and are no longer resonant with those in the center: 
a QH island representing a quantum magnetic bottle emerges in the central TINC region between magnetic tunnel barriers close to the leads.

To confirm the above qualitative statements numerically, we consider the setup of Fig.~\ref{fig_CB}(a).
%\textcolor{violet}{, where $\sigma\!=\!10\,\mu$m$^{-1}$}. 
For convenience we work with the coaxial coordinate $z$ instead of $s$. 
The radius $R_\sigma(z)$ is now characterized by a smoothing parameter $\sigma$
introducing surface curvature, ${\rm d}^2R_\sigma/{\rm d}z^2 \neq 0$. 
The ideal TINC from Fig.~\ref{fig_cone} is recovered for $\sigma\to\infty$
\cite{SuppMat}, see Fig.~\ref{fig_CB}(a).
Orange lines in Fig.~\ref{fig_CB}(b) mark the bound state energies of the potential wedges $|V_l(z)|$,
obtained by solving the generalized version of Eq.~\eqref{1D_Dirac}.  QH states with $n\!>\!0$ close to the leads are strongly lowered in energy as compared to the central ones, which still form the (disorder-broadened) LLs expected in the limit $\sigma\!\rightarrow\!\infty$.  
Effective magnetic confinement requires the smooth TINC border region 
to be larger than the (local) magnetic length $l_B$, so that off-resonant QH states can form, corresponding to $B$-field strengths of order 5-10 T for the smooth TINC considered.  
Larger TINCs or smoother border regions allow for weaker magnetic fields.
Notably, the $n\!=\!0$ LL is unaffected by the magnetic flux modulation, 
a distinct feature of Dirac electrons.
This leads to an anomalous coexistence of transport mechanisms: $n=0$ electrons stay delocalized and the associated conductance resonant, 
while potential barriers close to $z_0$, $z_1$ confine $n\neq 0$ electrons, 
whose transport properties are then dominated by Coulomb repulsion.
Indeed, when the Fermi energy aligns with an $n\neq 0$ LL, oscillations of Coulomb blockade type \cite{beenakker1991b} should be observed by varying the gate voltage $V_g$ applied to the central TINC region.

An exact description of blockade oscillations in strong magnetic fields is a highly non-trivial many-body problem that, in the case of semiconductors, is often solved self-consistently \cite{Meirav1996}.
However, the general features of the addition spectrum can usually be obtained from the constant interaction model, characterized by a constant charging energy $E_C$, and the single-particle spectrum of the island \cite{Meirav1996}.
Hence, we stay within such a model and discuss the expected consequences.
Figure~\ref{fig_CB}(c) shows the addition spectrum of the inner LLs as function of $B$ assuming homogeneous gating and $E_C\!=\!1\,$meV (this value is justified below). The $n=0$ LL is open to the leads and thus not split by interaction effects. 
Conductance peaks belonging to the $n=1$ LL are constantly spaced in gate voltage. % and disperse $\propto\sqrt{B}$.
The large gap in the upper panel is the Landau gap between the first and second LLs. For increasing $B$, additional QH states join the $n=1$ LL, adding to its degeneracy and shifting upwards the bottom of the $n=2$ LL in steps of $E_C$.
If the coupling $\Gamma_n^{l,r}$ to the left $(l)$ and right $(r)$ leads is roughly the same for all QH states within the n-th LL,
one expects conductance peaks $G_n \propto \Gamma_n^l\Gamma_n^r/(\Gamma_n^l+\Gamma_n^r)$ \cite{beenakker1991b} .
%Note that $\Gamma < E_C$ is needed to resolve such an addition spectrum experimentally.  
Besides the anomalous $n\!=\!0$ conductance peak, further differences between standard Coulomb blockade \cite{Meirav1996} 
and the present scenario are evident: 
First, the blockade conductance is determined by the Dirac spectrum $\propto\sqrt{B}$, 
not the one $\propto B$ of trivial electrons;
Second, the island-lead coupling -- and thus the conductance peaks height and shape -- is determined by the tunable magnetic barrier 
$\Gamma_n^{l,r}(B)$.  Its precise characterization is a non-trivial, setup-dependent problem.
Finally, the $n\neq 0$ islands are neither featureless resonant levels as in the simple quantum dots,
nor do they host QH edge states running along their perimeter \cite{Meirav1996}.
In fact, transport across each island is akin to that through a peculiar disordered system
whose transmission increases with increasing disorder [see Fig.~\ref{fig_cone}(b)].
Thus, our setting might host a competition between magnetically-induced blockade physics 
and a suitable generalization of (longitudinal) magnetotransport physics for 2D, disordered systems \cite{dmitriev2012, schirmacher2015}, which is unexplored ground.
Its study appears as a very interesting problem, albeit clearly beyond the scope of this work.
%\textcolor{red}{(Refs. to add: Dmitriev et al., RMP 2012, Schirmacher et al. PRL 2015.  
%\textcolor{violet}{
%Secondly, the $n\neq 0$ islands are neither featureless resonant levels as in the simplest quantum dots,
%nor do they host quantum Hall edge states running along their perimeter \cite{Meirav1996}.  
%In fact, in the absence of interactions transport across each island is of diffusive type,
%akin to that through a disordered network connecting the outer and inner edges of a Corbino disk
%-- this is a peculiar network, where hopping {\it increases} with disorder strength, cf.~Fig.~\ref{fig_cone}(a). }

\textit{Experimental realization} -- 
The parameters used are within experimental reach: HgTe-based 3DTI tubes~\cite{ziegler2018} or
core-shell nanowires~\cite{kessel2017} with spatially varying cross sections have already been built for transport experiments.
Crucially, our conclusions do not require the TINC to have a truly circular cross section: Surface deformations lift the degeneracies of the QH states within each LL,
but as long as this splitting is smaller than the broadening due to, {\it e.g}.~temperature or disorder, 
no qualitative conductance change is expected. 
Moreover, the leads are not required to have a cylindrical shape. 
To achieve our results, it is merely important that there are enough lead modes available that couple to the states on the TINC.
The charging energy $E_C$ depends on the setup size, geometry and materials, including the dielectrics,
and can thus be tuned in a broad range. For a HgTe-based TINC with parameters 
from Ref.~\cite{ziegler2018}, $E_C$ is of the order of $1\,$meV.
Homogeneous gating can be achieved with established techniques 
%{\it e.~g.~}``surrounding-gate architecture'' or ``lateral wrap-gate'' 
\cite{Storm2012,Royo2017}.

\textit{Conclusions} --
Shaped TINWs represent a broad class of mesoscopic junctions realizable with current technology
which allow to explore radically different magnetotransport regimes.  The latter result from the interplay
between magnetic confinement of Dirac electrons, disorder and interactions.
They are all accessible in TINCs by tuning a homogeneous magnetic field,
and orienting it in space.  TINCs and junctions thereof could act as building blocks
of more complex 3D mesoscopic Dirac electron setups, where, \textit{e.g.}, exotic aspects of QH physics on curved surfaces can be studied
\cite{lee2009, can2016}, or of TI-superconducting systems where Majoranas are looked for \cite{hasan2010,cook2011,manousakis2017}.

{\it Acknowledgements} -- We thank Andrea Donarini, Benjamin Geiger, Milena Grifoni, Dmitry Polyakov, and Alex Kamenev for useful discussions. 
This work was funded by the Deutsche Forschungsgemeinschaft (DFG, German Research Foundation) -- Project-ID 314695032 -- SFB 1277 (project A07 and B07), and within Priority Programme SPP 1666 ``Topological Insulators'' (project Ri681-12/2). Support by the Elitenetzwerk Bayern Doktorandenkolleg ``Topological Insulators'' is also acknowledged.

%\cite{Susskind1977}, \cite{Stacey1982}, \cite{Habib2016}

%\bibliographystyle{apsrev4-2}
%\bibliography{ultramagnetocone_biblio}

%\input{cone_27092019_FINAL.bbl}

%%%%%%%%%%%%%%%%%%%%%%%%%%%%%%%%%%%%%%%%%%%%%%%%%%%%%%%%%%%%%%%%%%%%%%%%%%%%%%%%%%%%%%
%%%		Bibliography
%%%%%%%%%%%%%%%%%%%%%%%%%%%%%%%%%%%%%%%%%%%%%%%%%%%%%%%%%%%%%%%%%%%%%%%%%%%%%%%%%%%%%%


%%%%%%%%%%%%%%%%%%%%%%%%%%%%%%%%%%%%%%%%%%%%%%%%%%%%%%%%%%%%%%%%%%%%%%%%%%%%%%%%
\cleardoublepage

%%%%%%%%%%%%%%%%%%%%%%%%%%%%%%%%%%%%%%%%%%%%%%%%%%%%%%%%%%%%%%%%%%%%%%%%%%%%%%%%
%%%		Supplemental material 
%%%%%%%%%%%%%%%%%%%%%%%%%%%%%%%%%%%%%%%%%%%%%%%%%%%%%%%%%%%%%%%%%%%%%%%%%%%%%%%%

%\input{SuppMat_neo.tex}

\end{document}

%\onecolumngrid
% 
% \begin{center}
%   \textbf{\Large{Supplemental material to the paper \\[1ex]
%       ``Magnetoconductance, Quantum Hall Effect, and Coulomb Blockade \\	
%          in Topological Insulator Nanocones''}}
% \end{center}
% \begin{center}
%   Raphael Kozlovsky, Ansgar Graf, Denis Kochan, Klaus Richter, Cosimo Gorini\\
%   \emph{\small Institut f\"ur Theoretische Physik, Universit\"at
%     Regensburg, D-93040 Regensburg, Germany }
% \end{center}
% \twocolumngrid

\title{\Large{Supplemental material to the paper \\[1ex]
		       ``Magnetoconductance, Quantum Hall Effect, and Coulomb Blockade \\	
		          in Topological Insulator Nanocones''}}
\author{Raphael Kozlovsky, Ansgar Graf, Denis Kochan, Klaus Richter, Cosimo Gorini}
%\email[]{cosimo.gorini@physik.uni-regensburg.de}
\affiliation{Institut f{\"u}r Theoretische Physik, Universit{\"a}t Regensburg, 
	93040 Regensburg, Germany}

\date{\today}
\maketitle

\setcounter{equation}{6}

\setcounter{figure}{4}

\section{Numerical implementation}

\comment{
\textit{Kwant} takes as an input tight-binding Hamiltonians of nanostructures and applies the so-called wave function approach to compute their electronic transport properties using the Landauer-B\"uttiker formalism  \kwant. Tight-binding Hamiltonians are passed to kwant as ``conventional'' Hermitian matrices, i.e. finite dimensional matrices which fulfill $(H^*)^T=H$. However, discretizing the continuum Hamiltonian Eq.~\hamiltonian ~leads to a tight-binding representation which does not fulfill this condition. Let us clarify that the reason for this is the non-trivial (non-constant, i.e. coordinate dependent) volume form by using a proof by contradiction:
Consider, for simplicity, a one-dimensional system with $\mrm{d}V = g(q)dq$. On the lattice, we use the shorthand notation $g_i \equiv g(q_i)$ as well as $\Psi_i \equiv \Psi(q_i)$, where $i$ labels the lattice sites. Then, if we assume that $(H^*)^T=H$, we have
\begin{align}
\left< \Phi | H \Psi \right>  \equiv  \sum_{ij} g_i \Phi_i^* H_{ij} \Psi_i  = \sum_{ij} g_i \left(H_{ji} \Phi_i\right)^* \Psi_j. \\
\end{align}
The Hermiticity of $H$ (defined as $\left< \Phi | H \Psi \right> = \left< H \Phi |  \Psi \right>$), then implies that 
\begin{align}
\sum_{ij} g_i \left(H_{ji} \Phi_i\right)^* \Psi_j  =  \sum_{ij} g_j \left(H_{ji} \Phi_i\right)^* \Psi_j.
\end{align}
In general, this is only fulfilled if $g(q) = \mathrm{const}$ and thus the assumption $(H^*)^T=H$ cannot hold for a non-trivial volume form.

Luckily, we can make the volume form trivial by applying the local transformation (conformal mapping?)  $ \widetilde{\Psi} = \sqrt{g} \Psi$ and $\tilde{H} = \sqrt{g} H  \frac{1}{\sqrt{g}}$, which leads to a scalar product $\left< \widetilde{\Phi} | \widetilde{\Psi} \right> = \int \! \mathrm{d} q  \, \widetilde{\Phi}^*(q) \widetilde{\Psi}(q)$.
}

% 2nd try
%Our quantum transport simulations are based on the tight-binding software package \textit{kwant}.
\textit{Kwant} takes as an input tight-binding Hamiltonians of nanostructures and applies the so-called wave function approach to compute their electronic transport properties using the Landauer-B\"uttiker formalism  \kwant.
Tight-binding Hamiltonians are passed to kwant as ``conventional'' Hermitian matrices, i.e. finite dimensional matrices which fulfill $(H^*)^T=H$. However, discretizing the continuum Hamiltonian Eq.~\hamiltonian ~leads to a tight-binding representation which does not fulfill this condition. Let us clarify that the reason for this is the non-trivial (non-constant, i.e. coordinate dependent) volume form. Consider, for simplicity, a one-dimensional system with real space coordinate $q$ and volume form $\mrm{d}V = g(q)dq$. On the lattice, we use the shorthand notation $g_i \equiv g(q_i)$ as well as $\Psi_i \equiv \Psi(q_i)$, where $i$ labels the lattice sites.
The condition for the Hermiticity of $H$, $\left< \Phi | H \Psi \right> = \left< H \Phi |  \Psi \right>$, is then given by
\begin{align}
	\sum_{ij} g_i \Phi_i^* H_{ij} \Psi_j = \sum_{ij} g_j \left(H_{ji} \Phi_i\right)^* \Psi_j
	\label{eq:hermiticity_on_lattice}
\end{align}
on the lattice. From Eq. \ref{eq:hermiticity_on_lattice}, it is, in general, only possible to deduce that  $(H^*)^T=H$ if $g(q) = \mathrm{const}$. Hence, we use the local transformation 
\begin{align}
	\label{eq:volume_form_trafo}
   H &\rightarrow \tilde{H} = \sqrt{g(q)}H (1/\sqrt{g(q)}), \\
   \Psi &\rightarrow \tilde{\Psi} = \sqrt{g(q)} \Psi
\end{align}
which makes the volume form trivial. This can be easily seen by considering the scalar product
\begin{align}
	\left< \Phi | H \Psi \right> &= \int_{-\infty}^\infty \! \mathrm{d} q \, g \Phi^* H \Psi \\
	&= \int_{-\infty}^\infty \! \mathrm{d} q \left(\sqrt{g} \Phi^*\right)  \left(\sqrt{g}H \frac{1}{\sqrt{g}}\right) \left(\sqrt{g} \Psi  \right) \\
	\label{eq:scalar_procuct_transformed}
	&= \int_{-\infty}^\infty \! \mathrm{d} q \, \tilde{\Phi}^* \tilde{H} \tilde{\Psi}. 
\end{align}
Note that in Eq. \ref{eq:scalar_procuct_transformed}, $g$ is fully absorbed in the wave functions and the Hamiltonian, the volume form is trivial.
Hence, using the transformation Eq. \ref{eq:volume_form_trafo} yields a tight-binding Hamiltonian $\tilde{H}$ which fulfills $(\tilde{H}^*)^T=\tilde{H}$. 

% numerical details
We implement the magnetic fields via usual Peierls phases and consider
standard Gaussian-correlated disorder,
\begin{equation}
\label{eq_disorder}
\langle V(\br) V(\br')\rangle = K\frac{\hbar v_F}{2\pi\xi^2} e^{-|\br-\br'|^2/2\xi^2} \, , 
\end{equation}
with disorder strength $K$ and correlation length $\xi$. For our simulations we use $K=0.1$ and $\xi=7$ nm.
%In order to compute the conductance, highly-doped semi-infinite cylindrical leads (with radii $R(s_0)$ and $R(s_1)$)
%are attached to the conical scattering region.
To avoid the fermion doubling problem \susskindstacey,
we use a conventional Wilson mass term.
For a recent discussion see Ref. \habib.  Note that a Wilson mass affects the specifics of
	the spin-momentum texture. Such specifics are, however, irrelevant for our charge transport problem.

\section{Effective mass potential for the TI nanocone in weak coaxial magnetic field}
\begin{figure*}
	\includegraphics[width=1.6\columnwidth]{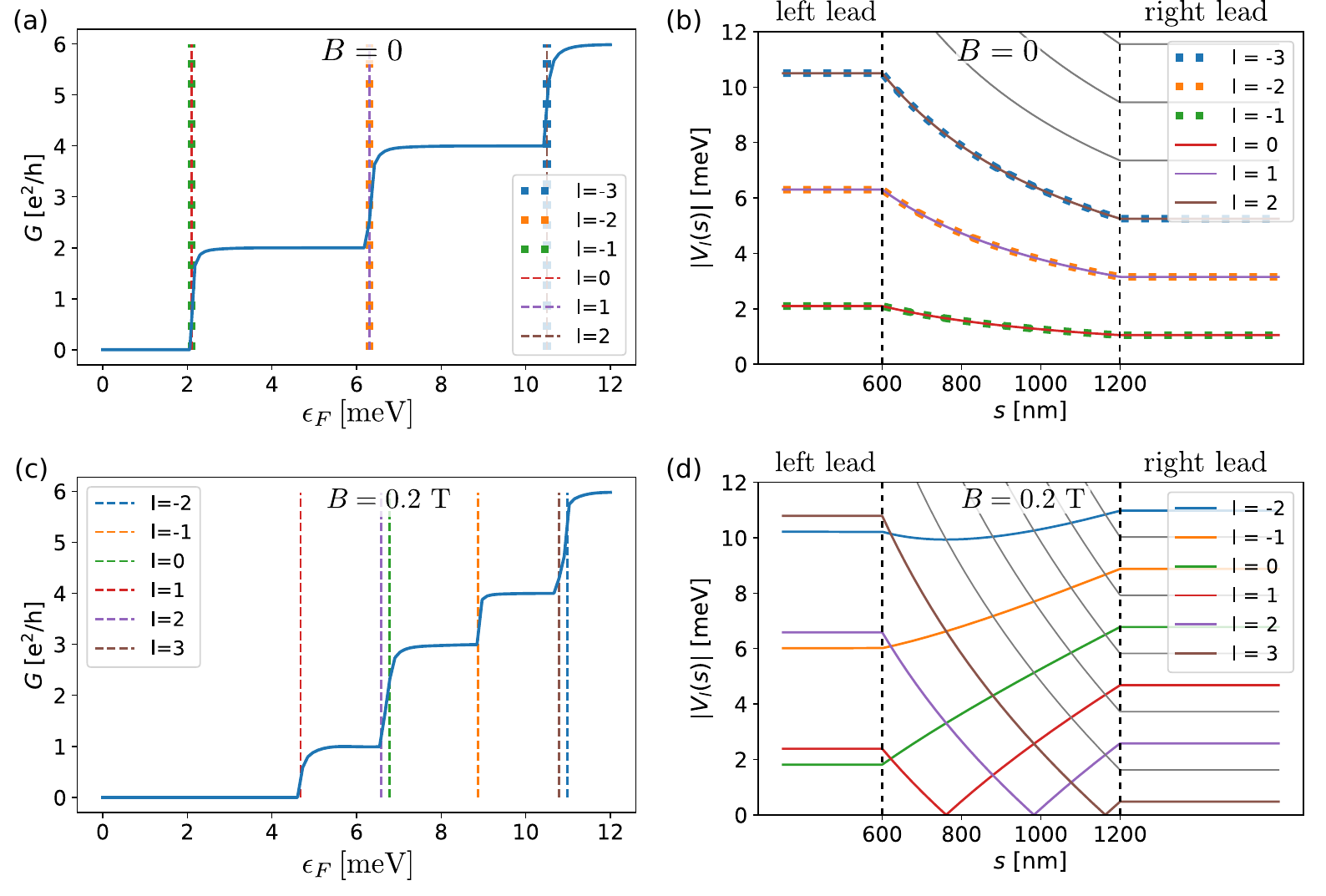}
	\caption{Conductance as a function of Fermi energy (left panels) and effective potentials (right panels) for a clean TI nanocone with leads, see Fig.~1(c). 
		Panels (a) and (b) correspond to $B=0$, panels (c) and (d) to a coaxial magnetic field $B = 0.2$ T. 
		Steps in the conductance appear whenever the Fermi energy $\epsilon_F$ surpasses one of the effective potentials $|V_l(s,B)|$ labeled by $l$ (corresponding energies are marked by vertical lines). For the conductance simulations we used \textit{kwant} and a Fermi velocity of $v_F=\SI{5e5}{m/s}$.
		%The effective potentials of the TINC are extended in the lead regions (left and right of the vertical black line). 
		%As soon as the Fermi energy is large enough such that modes with the same $l$-value exist in both leads, a transmission channel opens yielding a step in the conductance (see main text). The corresponding Fermi energies are marked with vertical lines in panels a) and c). 
		%The parameters of the cone are the same as those in Fig.~\ref{fig:effective_potential}.
		%In order to avoid Fermion doubling, the conductance was divided by four, which is justified since scattering (and thus inter-valley scattering) is absent without disorder.	
	}
	\label{fig:conductance_B=0_and_weak_B}
\end{figure*}
In order to develop a feeling of how the effective potential $|V_l(s,B)|$ defined in Eq.~\effPot~affects transport, we consider the conductance through a clean TINC (with the same geometry as in Fig.~\conductanceCone) for zero and weak coaxial magnetic field without doping in the leads.

The conductance for $B=0$ as a function of the Fermi energy $\epsilon_F$, depicted in Fig.~\ref{fig:conductance_B=0_and_weak_B} (a), shows equidistant conductance steps of size $2 e^2/h$. Their position and height can be explained with the corresponding effective potentials, which are plotted in Fig.~\ref{fig:conductance_B=0_and_weak_B}(b) for all $l$-values which are relevant in the depicted energy range. 
The central region in between the black vertical lines (600 nm $\leq s \leq $ 1200 nm) corresponds to the truncated cone, while the regions $s < 600\,$nm and $s >1200\,$nm correspond to straight leads. The leads are modeled by semi-infinite cylindrical nanowires, thus $|V_l(s,B)|$ is constant. Note that for $B=0$, the effective potential $|V_l(s)| \propto |l+0.5|/R(s)$ is purely determined by the finite size confinement energy, hence the spacing in the left lead (smaller circumference), is larger than in the right lead (larger circumference).
Modes which enter the scattering region (the truncated cone) via the leads are described by a well defined orbital angular momentum quantum number $l$, and each of them feels the effective potential $|V_l(s)|$.
As soon as the Fermi energy surpasses the effective potential for a given $l$, the corresponding mode has enough energy to traverse the TINC and a step in the conductance appears. Since there is no disorder and modes are twofold degenerate in orbital momentum, the conductance is quantized in units of $2e^2/h$.

%As soon as the Fermi energy in the lead surpasses the effective potential, modes start to enter the cone.
%In the right lead, this is the case at $\approx 1$ meV (for $l=-1$ and $l=0$). However, at this Fermi energy, there are no modes available in the left lead to scatter into. 
%Hence, the conductance stays zero until $\approx 2$ meV where the corresponding modes open in the left lead. This is where the first conductance step appears in Fig.~\ref{fig:conductance_B=0_and_weak_B} a). For energies larger than $\approx 2$ meV, a mode which enters with $l=0$ (or $l=-1$) from the right lead gains angular momentum while traversing the cone due to the increasing effective potential, and thereby looses longitudinal momentum. 
%Since there is no disorder, it reaches the left lead without scattering and the conductance is quantized in units of $e^2/h$.
%Due to the degeneracy of the modes for zero magnetic field the conductance steps are of the size $2 e^2/h$. 

%The energies where modes open in the left lead (and are thus open in both leads) are marked by vertical lines and colored according to their $l$-values in Fig.~\ref{fig:conductance_B=0_and_weak_B} a) [same color code as in Fig.~\ref{fig:conductance_B=0_and_weak_B} b)]. Note that the conductance steps at higher energies can be explained in the same way, and that the equidistant spacing between the conductance steps originates from the equidistant spacing of the subbands in the leads.
%

In Fig.~\ref{fig:conductance_B=0_and_weak_B} (c) and Fig.~\ref{fig:conductance_B=0_and_weak_B} (d), the conductance and the effective potential for a coaxial magnetic field of $B=0.2\,$T is shown. 
For low energies $\lesssim$ 2 meV, the characteristic effective potential ``wedges'', ubiquitous in Fig.~\conductanceCone (b) (strong magnetic field $B=2\,$T), start to emerge.
In the conductance, a more complicated step sequence appears with step sizes of $e^2/h$ and $2 e^2/h$ since orbital angular momentum degeneracy is broken due to the magnetic flux. 
The first step in the conductance, which is of size $e^2/h$, appears at $\approx 4.75$ meV which corresponds to the opening of the $l=1$ mode [red curve in (d)].
%In Fig.~\ref{fig:conductance_B=0_and_weak_B} c), we mark the energies where modes for a given $l$ appear in both leads and thus lead to conductance steps.
Modes with $l=0$ (green curve) and $l=2$ (purple curve) appear almost at the same energy around $6.75$ meV. %However, the purple mode opens in the left lead, whereas the green mode opens in the right lead. \\
The proximity of those two mode openings leads to a step size in the conductance of $2e^2/h$. 

Note that due to rotational symmetry, modes cannot change their orbital angular momentum quantum number $l$. %\textit{i.e.}~they cannot enter, for instance, with $l=0$ on the left side and exit with $l=3$ (brown curve) on the right side. 
However, as soon as rotational symmetry is broken, for instance, by disorder, modes couple. This coupling forms the basis of the resonant transport through Dirac LLs (consisting of bound states of effective potential wedges) discussed in the main manuscript.

\vspace{0.4cm}
\section{Smoothed TI nanocone}
For the description of a smoothed TINC we choose to work with the coaxial coordinate $z$. We approximate the Heaviside step function by $\Theta_\sigma(z-z')=\frac{1}{2}+\frac{1}{\pi}\arctan[\sigma (z-z')]$, $\lim\limits_{\sigma \rightarrow \infty}{\Theta_\sigma}(z-z')=\Theta(z-z')$. The radius of a junction that starts with a cylinder of radius $R_0$ at $z=-\infty$ and becomes a cylinder of radius $R_1$ at $z=\infty$ with an intermediate smoothed TINC can then be written as
\begin{equation}
\begin{aligned}
R_\sigma(z)=& R_0+(R_1-R_0)\Theta_\sigma(z-z_1)\\
&+\mathcal{S}(z-z_0)[\Theta_\sigma(z-z_0)-\Theta_\sigma(z-z_1)],
\label{radius}
\end{aligned}
\end{equation}
where $\mathcal{S}=(R_1-R_0)/(z_1-z_0)$ is the slope of a TINC with initial radius $R_0$, final radius $R_1$ and length $z_1-z_0$. 
The input parameters chosen for Fig. 4 are such that in the limit $\sigma\mapsto\infty$ the function (\ref{radius}) agrees with the TINC studied in Fig. 3. Thus $R_0=C_\text{min}/(2\pi)=78.3\,$nm, $R_1=2R_0=C_\text{max}/(2\pi)\approx156.6\,$nm, $z_0=0$ and $z_1=594.7\,$nm.

%
% Fill the last column with empty space, to avoid a stretched colum
\vspace*{\fill}